\documentclass[twocolumn,secnumarabic,amssymb,amsmath,showpacs,nobibnotes, aps, prb]{revtex4-1}
\usepackage{graphicx}% Include figure files
\usepackage{amsmath}
\usepackage{color}

\begin{document}
\title{2-dimensional hyperbolic medium for electrons and photons based on the array of tunnel-coupled graphene nanoribbons }% Force line breaks with \\
%\thanks{Footnote to title of article.}
\author{Iurii Trushkov$^{1}$}
\author{Ivan Iorsh$^{1}$}
%\author{Pavel Belov????$^{1}$}
\affiliation{$^1$National Research University of Information Technologies,
Mechanics and Optics (ITMO), St.~Petersburg 197101, Russia}

\begin{abstract}
We study the electronic band structure and optical conductivity of an array of tunnel-coupled array of graphene nanoribbons. We show that due to the coupling of electronic edge states for the zigzag nanoribbon structure, the Fermi surface can become a hyperbola similarly to the case of the layered metal-dielectric structures, where the hyperbolic isofrequency contours originate from the coupling of localized surface plasmon polaritons. Moreover, we show that for both types of the ribbon edge, the optical response of the structure can be characterized by a uniaxial conductivity tensor, having principal components of the different signs. Therefore, the tunnel-coupled nanoribbon array  can be regarded as a tunable hyperbolic metasurface.
\end{abstract}

\maketitle
             
\section{Introduction}
Graphene is a two-dimensional carbon lattice  that exhibits a wide range of unique electronic properties such as linear dispersion of charge carriers, Berry phase, and room-temperature quantum Hall effect~\cite{Rev1,Rev3}. Beyond the purely electronic properties, graphene has attracted a lot of attention in the field of nanophotonics, since it is both transparent and exhibits strong light-matter interaction~\cite{GR_Ph1,GR_Ph2}. Moreover, graphene supports the surface electromagnetic waves, analogous to surface plasmon polaritons in metallic and semiconductor films. The advantage of the graphene plasmon-polaritons over their metallic and semiconductor counterparts is that they can be efficiently controlled with the external gate voltage~\cite{Koppens_exp,Basovexp}. The attractive plasmonic properties of graphene nanostructures  led to the emerging of the field of graphene nanoplasmonics~\cite{GraphPlasm_Grigorenko,GraphPlasm_Abajo}.

One of the topics in graphene plasmonics is the study of graphene \textit{metasurfaces}, the arrays of graphene nanoislands which may exhibit the optical properties which differ significantly from those of the two-dimensional graphene sheets~\cite{GraphSRR, GrapheneMM1,GrapheneMM2,GrapheneMM3, GrapheneSoukoulis, Engheta_Science}. For example, it has been shown that a two-dimensional array of graphene nanoislands can play the role of a perfect absorber for electromagnetic waves~\cite{Perfect_Absorption}. In the studies of graphene-based metasurfaces it is commonly assumed that the conductivity of individual graphene elements coincides with the conductivity of two-dimensional graphene. However, it is known from  numerous studies that the graphene nanopatterning can substantially alter the electronic band structure of graphene~\cite{GrNS_BD1,GrNS_BD2}. Specifically, in Ref.~\cite{GrNS_BD1} it was shown that one can open and efficiently control the width of the band gap by nanopatterning graphene. Altering the electronic band structure should lead to the modification of the AC conductivity and thus to the modification of the optical properties of the nanopatterned graphene.
\begin{figure}[!h]
\centerline{\includegraphics[width = 1.0\columnwidth]{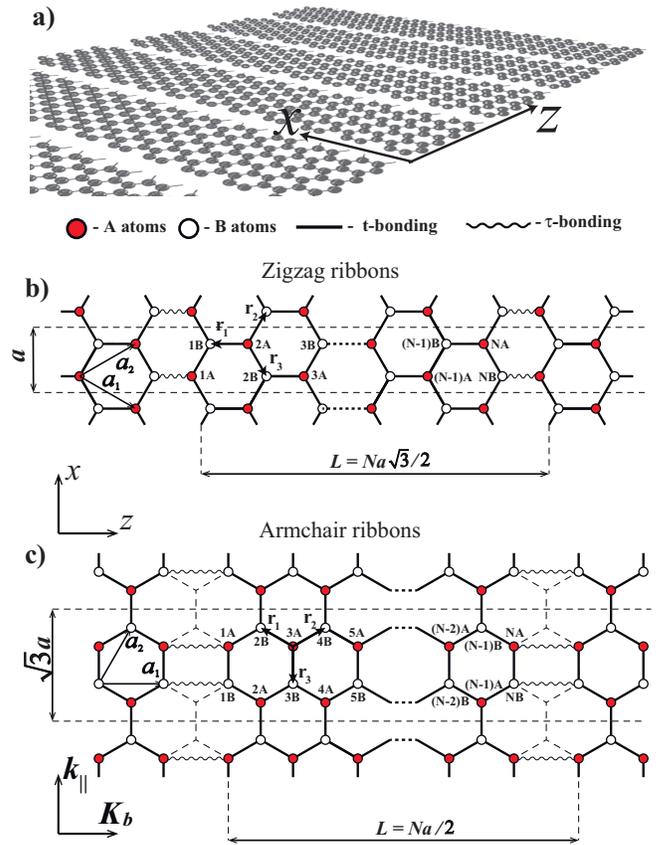}}
\caption{(Color online) (a) Geometry of the structure (b,c) Structure  the array of tunnel coupled array of zigzag (b) or armchair (b) graphene nanoribbons.}
\label{fig1}
\end{figure} In Ref.~\cite{Abajo1} it was theoretically shown that the quantum confinement effects start play the substantial role in the optical response of the patterned graphene if the characteristic size of the nanostructure elements becomes smaller than 10 nm.
Technically, the introduction of the superlattice to the graphene sheet at these scales can be realized  by either etching the graphene sheet~\cite{etching}, introducing the periodic strain~\cite{Corrugation1,Corrugation2,Metallic_Substrate_Mask} or by introducing the periodic gate voltage~\cite{Superlattice_2_technology,Superlattice_3_technology}. Moreover, in Ref.~\cite{Self-organized_GMS} the possibility of molecular assembly of the arrays of several-atoms-thick nanoribbons was shown.

In our work we are studying the structures shown in Fig.\eqref{fig1}, which are essentially the arrays of tunnel-coupled graphene nanoribbons. We demonstrate that the optical properties of these structures can be tailored by engineering the electronic band structure, and that such a system can be regarded as a uniaxial anisotropic medium both for electrons and plasmon-polaritons. The optical properties of the graphene nanoribbon arrays have been studied in a number of papers. Specifically, in~\cite{NMR} it has been shown that a nanoribbon array can play a role of an efficient polarizers for the THz radiation due to the huge absorption anisotropy. However, in all of the papers the tunneling between the individual ribbons was not accounted for. The allowing for the electron hopping between the ribbons results in a number of physical effects such as emergence of hyperbolic electronic bands from the edge states in the zigzag ribbon case and overlap of electron and hole bands in the armchair edge case which will be discussed further in detail.

The remainder of the paper is organized as follows. In section~\ref{sec1} we show the results on the electronic band structure and Fermi surface topology of the arrays of zigzag and armchair graphene nanoribbons. Section~\ref{sec2} presents  the results on the conductivity spectra obtained with the Kubo formula. The conclusion of the paper is given in section~\ref{conc}. Finally, appendices~\ref{sec3} and~\ref{sec4} provide the detailed formalism for the calculation of the band structure and the conductivity, respectively.

\section{Band structure calculation\label{sec1}}
For both types of the ribbons edge the Hamiltonian can be written as~\cite{Theory_tight_binding}:
\begin{align}
\hat{H}=&-t\displaystyle\sum_{nearest} (\hat{a}^{\dag}_{\alpha} \hat{b}_{\beta}+ \hat{b}^{\dag}_{\beta} \hat{a}_{\alpha})-t^{'}\displaystyle\sum_{\substack{next \\ nearest}} (\hat{a}^{\dag}_{\alpha} \hat{a}_{\beta}+ \hat{b}^{\dag}_{\alpha} \hat{b}_{\beta}) - \nonumber \\
-&\tau\displaystyle\sum_{\substack{inter-\\ribbon}} (\hat{a}^{\dag}_{\alpha} \hat{b}_{\beta}+\hat{b}^{\dag}_{\beta} \hat{a}_{\alpha}), \label{Hamiltonian}
\end{align}
where the first term corresponds to the hopping of the electron between the nearest atoms with the characteristic hopping energy $t=2.8$~eV, the second term to the next-nearest hopping with $t^{\prime}=0.2$~eV, and the last term corresponds to the tunneling between the nearest ribbons with the tunneling energy $\tau$ which can be varied (see Fig.\ref{fig1}). Generally, $\tau$ should decrease exponentially with distance between the neighbouring ribbons $\tilde{a}$ $\tau\approx t\exp(-2(\tilde{a}-a)/a)$\cite{tunnel}, where $a$ is the interatomic distance. We note that as $\tilde{a}$ goes to infinity, the band structure and the conductivity should reduce to those of an individual nanoribbon. In the opposite limit $\tilde{a}=a$, the considered structure can either converge to the conventional 2d graphene case as in Fig.\ref{fig1}(b) or result in a more complicated structure different from the two-dimensional graphene as in Fig.\ref{fig1}(c). In our calculations we have used $\tilde{a}=\sqrt{3}a$ which corresponds to $\tau=t^{\prime}$. 

Indices $[\alpha,\beta]$ in the Eq.~\eqref{Hamiltonian} each consist of three integers: the first one  $i$  is  the number of the atom in the unit cell. The unit cells for the case of armchair and zigzag ribbons are shown with dashed lines in Figs.~\ref{fig1}(b,c). Each unit cell contains $N$ atoms of each sublattice. The remaining two indices $(m,n)$ correspond to the position of the unit cell in the directions of $K_b$ and $k$, respectively (cf. Figs~\ref{fig1}(b,c) for further details).
We then introduce Fourier transform of the annihilation and creation operators $a_{\alpha},a_{\alpha}^{\dagger}$:
\begin{align}
&a_{i,m,n}=\frac{1}{N_{c}^{1/2}}\sum_{k,K_b} a_{i}(k,K_b) e^{-i (K_b m L+k n a)}, \nonumber\label{afour}\\ 
&a^{\dag}_{i,m,n}=\frac{1}{N_{c}^{1/2}}\sum_{k,K_b} a^{\dag}_{i}(k,K_b) e^{i (K_b m L+k n a)}
\end{align}
where $N_c$ is the number of unit cells. The expressions for the operators $b$ are identical.
The wave functions of the structure are the linear combinations of the single atom eigenfunctions:
\begin{align}
\phi(k,K_b)=\displaystyle\sum_{i=1,N} \left[\mathcal{A}_i a_i^{\dagger}(k,K_b)|0\rangle +\mathcal{B}_i b_i^{\dagger}(k,K_b)|0\rangle \right],
\end{align}
and thus can be represented as a $2N$ vector of the expansion coefficients $\mathcal{A,B}$:
\begin{align}
\phi = \left(\mathcal{A}_1 \ldots \mathcal{A}_N,\mathcal{B}_1\ldots \mathcal{B}_N\right)^T.
\end{align}

If we substitute the expressions~\eqref{afour} to the Hamiltonian~\eqref{Hamiltonian} we obtain $H=\displaystyle\sum_{k,K_b}H(k,K_b)$, where $H(k,K_b)$ can be represented as a $2N\times 2N$ matrix in the basis of the single atom eigenfunctions:
\begin{align}
H=\begin{pmatrix} H_{AA} & H_{AB} \\ H_{BA} & H_{BB}  \end{pmatrix}. \label{hmat}
\end{align}
Explicit form of the matrix depends on the type of the ribbon edge and is presented in appendix~\ref{sec3} for both types of the edge, here we just note that due to the requirement of Hermiticity the condition $H_{BA}=H_{AB}^{\dagger}$ holds.

\begin{figure}[!h]
\centerline{\includegraphics[width = 1.0\columnwidth]{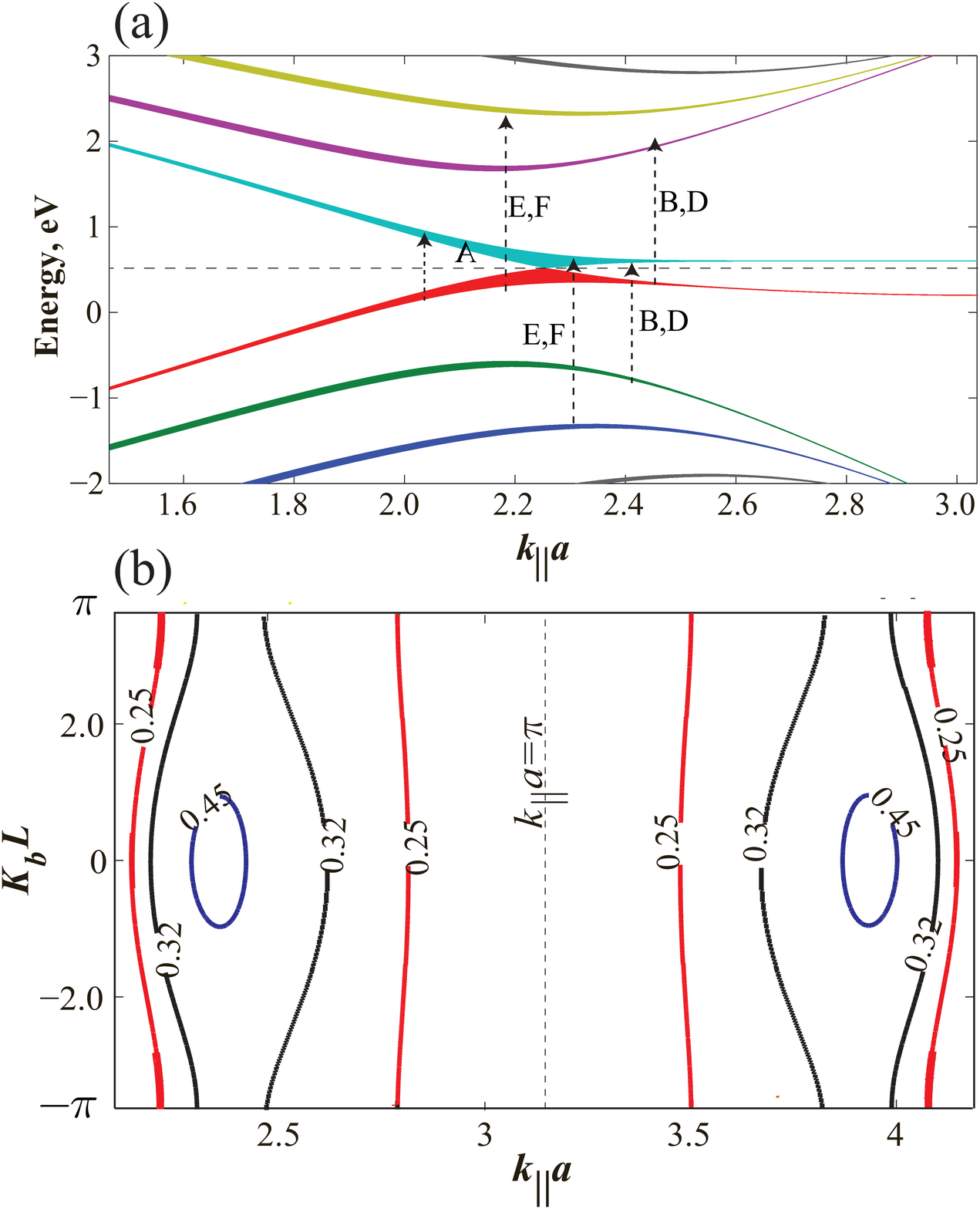}}
\caption{(Color online) Electronic band structure of the array of tunnel coupled zigzag nanoribbons. (b) Contour plots of the Fermi surfaces of the structure. The numbers at the curves define the corresponding electron energy.}
\label{fig2}
\end{figure}

\begin{figure}[!h]
\centerline{\includegraphics[width = 1.0\columnwidth]{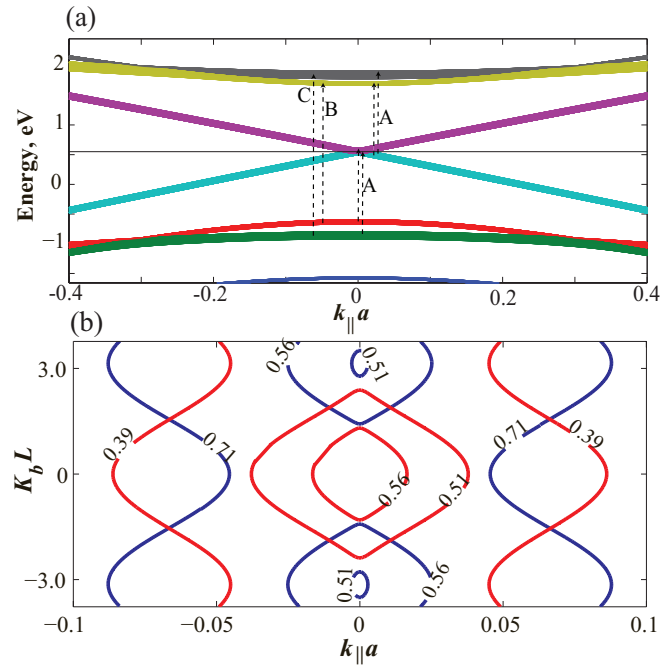}}
\caption{(Color online) Electronic band structure of the array of tunnel coupled armchair nanoribbons. (b) Contour plots of the Fermi surfaces of the structure. Blue (red) curves correspond to electron (hole) branches. The numbers on the curves define the corresponding electron energy. }
\label{fig3}
\end{figure}

The problem is  then reduced to the eigenvalue problem for a $2N\times 2N$ matrix $H(k,K_b)\phi=\epsilon(k,K_b)\phi$, which has been solved numerically in order to obtain the band structures and Fermi surfaces. For the case of zigzag ribbons $N=10$ and  for the case of armchair ribbons $N=11$. The results are presented in Figs.\ref{fig2},\ref{fig3}
We first consider the zigzag edge case shown in Fig.\ref{fig2}.  We first note,that due to the inclusion of the second nearest neighbours hopping, the electron-hole symmetry is broken and the electron and hole bands touch at the nonzero energy~\cite{RMP}. It can be seen that the individual branches of the single ribbon band diagram evolve to the bands. Moreover, the degeneracy of the two edge states is lifted and we observe two distinct bands formed by the tunnel coupled edge states. These bands resemble the plasmonic bands existing in the layered metal-dielectric structures~\cite{Orlov} and originating from the photon tunneling between individual surface plasmon polaritons propagating along the metal-dielectric interfaces. It is   these plasmonic bands that form the hyperbolic isofrequency contours and make the layered metal dielectric structures the most conventional realization of hyperbolic medium~\cite{Hyprev}. The coupled waveguide arrays have been show to exhibit hyperbolic media properties both for the case of electromagnetic~\cite{fishnet1,fishnet2} and acoustic~\cite{acoustic1} waves.  As can be seen in Fig.\ref{fig2}(b), where the energy contour plots are presented,  the similar situation takes place for the electrons in coupled zigzag ribbons. At the energy where the electron and hole bands touch (equal to $3t^{\prime}\approx 0.6$eV) the contour plots reduce to single points at the touching position. Then, as we shift to the area where the band formed by the edge states exists, we can observe the hyperbolic-like isofrequency contours. Thus, if the chemical potential is tuned such that the Fermi energy falls to the narrow region of these bands, the electrons at the Fermi surface would propagate in the effective hyperbolic media. As long as effectively ballistic propagation of electrons is considered, the negative refraction and the partial focusing of electron waves should be observed at the interface of the 2d graphene and patterned graphene structure just as for the case of electromagnetic waves~\cite{SmithPartial}. The hyperbolic isofrequency contours also lead to the enhanced density of electronic states as compared to the case of two-dimensional graphene at the same energy. We also note that at certain energies  the Fermi contour reduces to  parallel lines, and the two-dimensional structure effectively behaves as a one dimensional. As it is known~\cite{Pairing}, in the vicinity of such points the electron-electron interactions are substantially modified and the inclusion of the Hubbard repulsion to our tight-binding model is the subject of the future work. The advantage of the graphene-based structures is that the Fermi level can be fine-tuned with the external gate voltage, such that it falls to the region characterized by the hyperbolic isofrequency contour.

The band diagram and the Fermi surface  contour plots for the case of an armchair ribbons are shown in Fig.\ref{fig3}.
For the case of armchair ribbon, the Dirac cones exist only if the number of atoms in the unit cell $N=3m+2$, where $m$ is an integer~\cite{RMP}. In the considered structure $N=11$ and thus the Dirac cones are preserved. It can be seen  that coupling leads to the overlap of the electron and hole bands, and thus as can be seen in Fig.\ref{fig3}(b) at certain Fermi energies, the Fermi surface contains both electron and hole pockets. 

We now move to the calculation of the AC conductivity of the considered structures.

\section{Conductivity tensor.\label{sec2}} 
When the eigenvalues and eigenvectors are obtained, it is straightforward to compute the AC conductivity tensor of the structure using the Kubo formalism. The conductivity tensor is given by~\cite{Bruuc}
\begin{align}
\sigma_{\alpha \alpha}=\frac{\hbar}{iS} \sum_{s', s} \sum_{\substack{n,m\\k,K_{b}}} \frac
{\left( f[\varepsilon_{m}^{s'}]-f[\varepsilon_{n}^{s}] \right) \left|\langle \phi_{m}^{s'} |-ev_{\alpha}| \phi_{n}^{s} \rangle \right|^2}
{( \varepsilon_{m}^{s'}-\varepsilon_{n}^{s} ) ( \varepsilon_{m}^{s'}-\varepsilon_{n}^{s} +\hbar \omega+i\delta)}
\end{align}        
The velocity operator $\mathbf{v}$ is defined through  $-e\mathbf{v}/c=\partial H(\boldsymbol{A})/ \partial \boldsymbol{A}  $, where $ H(\boldsymbol{A})$
is derived from the tight-binding model via Peierls substitution:
\begin{align}
 t_{\boldsymbol{R},\boldsymbol{R'}} \to t_{\boldsymbol{R},\boldsymbol{R'}} e^{\frac{ie}{\hbar c}\int_{\boldsymbol{R}}^{\boldsymbol{R'}} \boldsymbol{A} d \boldsymbol{r} } =
 t_{\boldsymbol{R},\boldsymbol{R'}} e^{\frac{ie}{\hbar c} (\boldsymbol{R'}-\boldsymbol{R}) \boldsymbol{A}}.
\end{align}
The latter equality is made within the approximation that the vector potential is constant at the scale of single unit cell. The velocity operator then can be presented as an $2N\times 2N$ matrix of the form:
\begin{align}
-e\mathbf{v}=-e\begin{pmatrix}\mathbf{v}_{AA} & \mathbf{v}_{AB} \\ \mathbf{v}_{BA} & \mathbf{v}_{BB}\end{pmatrix}.
\end{align}
Due to the requirement of the hermiticity of the velocity operator, we set $\mathbf{v}_{AB}=\mathbf{v}_{BA}^{\dagger}$ and within the nearest neighbour approximation $\mathbf{v}_{AA}=\mathbf{v}_{BB}=0$.
The explicit form of $\mathbf{v}_{AB}$ depends on the type of the edge and is presented in the appendix~\ref{sec4}. The results are shown in Figs.\ref{fig4}(a,b). The temperature was set to $300$K and the chemical potential was equal to $3t^{\prime}$. The conductivity is normalized to the bare graphene conductivity $\sigma_0=e^2/h$
\begin{figure}[!h]
\centerline{\includegraphics[width = 1.0\columnwidth]{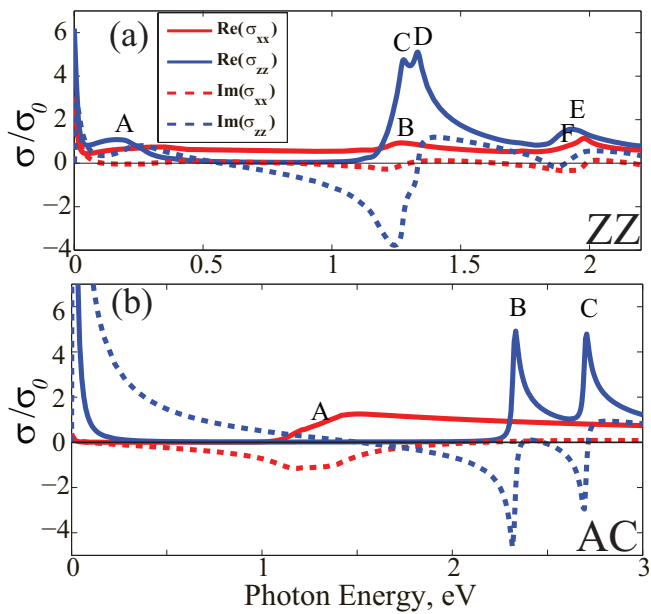}}
\caption{(Color online) Real and imaginary parts of the principal components of the conductivity tensor for the case of zigzag (a) and armchair (b) ribbon edge. Letters in (a) and (b) correspond to the transitions shown in Fig.\ref{fig2}(a) and Fig.\ref{fig3}(a) respectively.}
\label{fig4}
\end{figure} 
For the case of zigzag ribbons, we observe the strong absorption for the field polarized along the ribbons. This effect is due to the presence of the edge states and is also present in the conductivity spectra of individual zigzag ribbons~\cite{Single_Ribbon} and graphene nanodiscs~\cite{Single_Disc}. However, due to the lift of degeneracy between the edge states, an additional absorption channel between the two edge states bands appears, which results in the broad resonance of the absorption for the transverse-polarized electric field, depicted with letter A in Fig.\ref{fig4}(a). In the low frequency region, the conductivity has the Drude-like shape for both electric field polarizations. In the higher frequency region the conductivity is determined by the interband transitions shown in Fig.\ref{fig2}(a). Namely, the transitions between the second hole band and first electron band and first electron and second hole band result in the two distinct absorption peaks in $xx$ conductivity tensor component (marked with letters C,D) and one peak in $zz$ component (marked with B). Transitions between the first hole and third electron and first electron and third hole bands result in the higher frequency absorption peak for both conductivity components (marked with letters E,F).

For the case of armchair ribbons shown in Fig.~\ref{fig4}(b) it is worth noting that the interband absorption is almost fully suppressed for the transitions between the first electron and hole bands. This is quite different to the case of conventional graphene which is characterized by a broadband absorption caused by interband transitions. The suppression of interband absorption caused by the Fermi surface topology has been studied previously~\cite{suppr}. The detailed study of the interband absorption suppression in the considered geometry is the subject of the ongoing work. In the higher frequency region the absorption peaks are defined by the interband absorption between the higher electron and hole bands. We note that for the $xx$ component of the conductivity the peak in absorption is a step-like similar to the two-dimensional systems, and $zz$ conductivity has the peaks similar to the Van Hove singularities in one-dimensional systems. For the electromagnetic applications, it is important that in the certain frequency region the absorption is suppressed, and the imaginary parts of the conductivity tensor components have opposite signs, which suggests that we can realize a high quality anisotropic metasurface based on the considered structure.

\begin{figure}[!h]
\centerline{\includegraphics[width = 1.0\columnwidth]{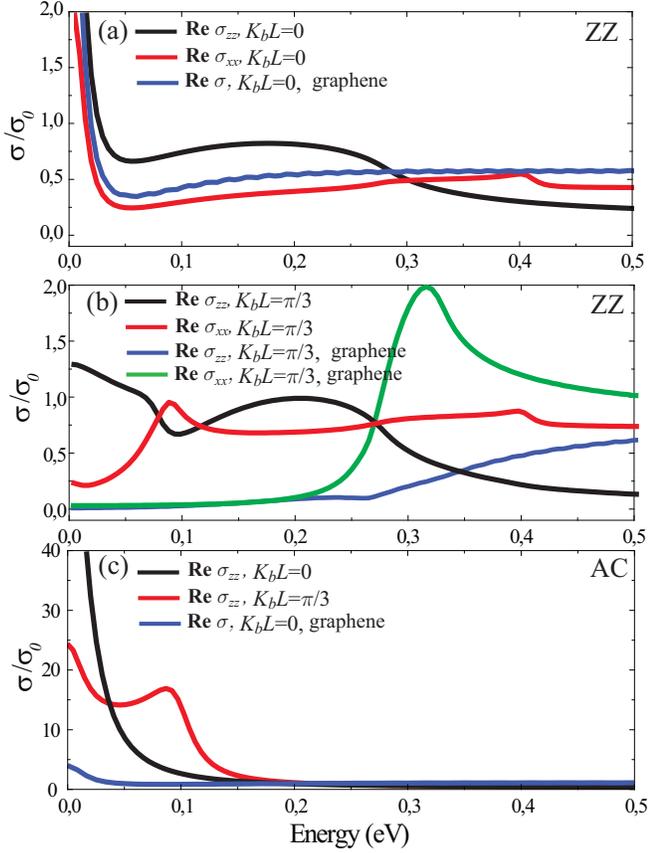}}
\caption{(Color online) Conductivity spectrum for the zigzag (a,b) and armchair (c) ribbons for different values of $K_bD$. The conductivity of two-dimensional graphene is shown for comparison.}
\label{fig5}
\end{figure}

We should also note, that the obtained conductivity tensors are nonlocal, i.e. the conductivity depends on the in-plane momenta of the electromagnetic wave . In Fig.~\ref{fig5} the real part of the conductivity of both zigzag and armchair ribbons is presented for different values of $K_b D$ and  compared to the conductivity of two-dimensional graphene. The substantial modification of the real part of the conductivity is observed due to the opening of new absorption channels via the indirect interband transitions.

\section{Conclusion and outlook.\label{conc}}
In this paper, we have analyzed the electronic band structure and conductivity spectrum of two-dimensional arrays of tunnel-coupled arrays of graphene nanoribbons. We have shown that both depend drastically on the type of the ribbon edges. Namely, for the case of zigzag edges the coupling of the electronic edge states lifts the degeneracy between the symmetric and antisymmetric electronic eigenmode formed by the edge states. Moreover, each of the individual modes evolves to a narrow band, defined by the hyperbolic isofrequency surfaces. We have thus demonstrated the simple yet interesting analogy between the considered structures and  multilayered metal-dielectric metamaterials, where the hyperbolic isofrequency surface for the electromagnetic field originates from the coupling between individual surface-plasmon polariton modes. For the case of armchair ribons, we have shown that the coupling leads to the overlap of electron and hole bands, resulting into the formation of the hole and electron pockets in the Fermi surface of these structures in the certain doping range. 

Furthermore, we have shown that these structures can be regarded as metasurfaces for the electromagnetic waves in the optical to mid-IR frequency range~\cite{Metasurface_Review}. Specifically, we have demonstrated that both types of the structures are characterized by different signs of imaginary part of conductivity tensor components together with small levels in some frequecy ranges. Recently, a phenomenological study of the electromagnetic spectrum of such metasurfaces has been performed, showing that these structures can support a special type of Dyakonov-like hybrid surface electromagnetic waves~\cite{arx1}.

The modification of the electronic band structures as well as electronic eigenfunction profiles in the nanostructured graphene should lead to the substantial modification of both electron-electron and electron-phonon interactions. Tailoring the electron-phonon scattering via the electron band engineering may lead to the substantial modification of the plasmon scattering rates and thus to the modification of the optical response of these structures~\cite{Pathways} Self-consistent calculation of the Coulomb self-energy as well as interactions-induced modification of the AC conductivity is the subject of future work. Availability to control the optical response of the graphene metasurfaces via the modification of the electronic band structure through the quantum confinement makes these structure a perspective platform for the novel optical to IR optoelectronic devices.

\appendix \section{Band structure calculation\label{sec3}}
Below is presented the explicit form of the Hamiltonian from Eq.~\eqref{hmat} for both types of the ribbons edge.

\textbf{Zigzag edge:}
\begin{align}
H_{AB}=-t\cdot
\begin{pmatrix}
\kappa & 0 & 0 & \ldots & 0& \frac{\tau}{t} \, e^{iK_{b}L}\\
1 & \kappa^{*} & 0  & \ldots & 0 & 0\\
0 & 1 & \kappa & \ldots & 0 & 0\\
\vdots & \vdots  & \vdots & \ddots & \vdots & \vdots \\
 0 & 0 & 0 & \ldots & \kappa & 0\\
 0 & 0 & 0 & \ldots &  1 & \kappa^{*}
\end{pmatrix},
\end{align}
\begin{align}
 H_{AA}=-2t'\cdot \begin{pmatrix}
\cos{ka} & \frac{\kappa}{2} & 0 &  \ldots  & 0\\
\frac{\kappa^{*}}{2} & \cos{ka} & \frac{\kappa^{*}}{2}  & \ldots &  0\\
0 & \frac{\kappa}{2} & \cos{ka} &  \ldots  & 0\\
 \vdots & \vdots & \vdots & \ddots & \vdots\\
 0 & 0 & 0 & \ldots  & \frac{\kappa}{2}\\
 0 & 0 & 0 & \ldots & \cos{ka}\end{pmatrix},
 \end{align}
 where $\kappa=1+e^{ika}$, $L=Na\sqrt{3}/2$ $H_{BA}=H_{AB}^{\dagger}, H_{BB}=H_{AA}^*$.

\textbf{Armchair edge:}
\begin{align}
H_{AB}=-t\cdot  
\begin{pmatrix}
1 & 1 & \ldots & 0 & 0  \\
1 & e^{i k \sqrt{3}a} &  \ldots & 0 & 0 \\
\vdots  & \vdots & \ddots & \vdots & \vdots \\
0 & 0 & \ldots & e^{i k \sqrt{3}a} & 1 \\
0 & 0 & \ldots &  1 & 1
\end{pmatrix},
\end{align}
\begin{align}
&H_{AA}= \nonumber \\
&-t'\cdot
\begin{pmatrix}
0 & \kappa^{*}  & 1  & \ldots & 0 & 0 &  \frac{\tau}{t'} e^{iK_{b}L}\\
\kappa& 0 & \kappa  & \ldots & 0 & 0 & 0\\
1 & \kappa^{*}   & \kappa^{*}  &  \ldots & 0 & 0 & 0\\
0 & 1 & \kappa  &  \ldots & 0 & 0 & 0\\
\vdots & \vdots & \vdots &  \ddots & \vdots & \vdots & \vdots \\
0 & 0 & 0  & \ldots & 0  & \kappa^{*} & 1\\
0 & 0 & 0  & \ldots & \kappa  & 0 & \kappa \\
 \frac{\tau}{t'} e^{-iK_{b}L} & 0 & 0  & \ldots & 1 &  \kappa^{*} & 0
\end{pmatrix},
\end{align}
\begin{align}
&H_{BB}= \nonumber \\
&-t'\cdot
\begin{pmatrix}
0 & \kappa  & 1  & \ldots & 0 & 0 &  \frac{\tau}{t'} e^{iK_{b}L}\\
\kappa^{*}& 0 & \kappa^{*}  & \ldots & 0 & 0 & 0\\
1 & \kappa   & \kappa  &  \ldots & 0 & 0 & 0\\
0 & 1 & \kappa^{*}  &  \ldots & 0 & 0 & 0\\
\vdots & \vdots & \vdots &  \ddots & \vdots & \vdots & \vdots \\
0 & 0 & 0  & \ldots & 0  & \kappa & 1\\
0 & 0 & 0  & \ldots & \kappa^{*}  & 0 & \kappa^{*} \\
 \frac{\tau}{t'} e^{-iK_{b}L} & 0 & 0  & \ldots & 1 &  \kappa & 0
\end{pmatrix},
\end{align}
where $\kappa=1+e^{ika\sqrt{3}}$,$L=Na/2$, $H_{BA}=H_{AB}^{\dagger}$.

\section{Calculation of the AC conductivity\label{sec4}}.
The expression for the conductivity operator for the case of zigzag edge is given by:
\begin{widetext}
\begin{align}
-e\boldsymbol{v}_{AB}=-\frac{ie}{\hbar}t\cdot
\begin{pmatrix}
\boldsymbol{r_{2}}+\boldsymbol{r_{3}}e^{ika} & 0 & 0 & \ldots & 0 & 0\\
\boldsymbol{r_{1}} & \boldsymbol{r_{3}}+\boldsymbol{r_{2}}e^{-ika} & 0 & \ldots & 0 & 0\\
0 & \boldsymbol{r_{1}} & \boldsymbol{r_{2}}+\boldsymbol{r_{3}}e^{ika}& \ldots & 0 & 0 \\
\vdots & \vdots & \vdots & \ddots & \vdots & \vdots \\
0 & 0 & 0 & \ldots & \boldsymbol{r_{2}}+\boldsymbol{r_{3}}e^{ika} & 0\\
0 & 0 & 0 & \ldots &  \boldsymbol{r_{1}} & \boldsymbol{r_{3}}+\boldsymbol{r_{2}}e^{-ika}
\end{pmatrix},
\end{align}
\end{widetext}
with $\mathbf{r_1}=a/\sqrt{3}(-1,0)$; $\mathbf{r_2}=a/\sqrt{3}(1/2,\sqrt{3}/2)$; $\mathbf{r_3}=a/\sqrt{3}(1/2,-\sqrt{3}/2)$,
and for the case of armchair edge is given by
\begin{align}
-e\boldsymbol{v}_{AB}=-\frac{ie}{\hbar}t\cdot
\begin{pmatrix}
\boldsymbol{r_{3}} & \boldsymbol{r_{2}} & 0 & \ldots & 0 & 0\\
\boldsymbol{r_{1}} & \boldsymbol{r_{3}}e^{ik\sqrt3 a} &\boldsymbol{r_{2}} & \ldots & 0 & 0\\
0 & \boldsymbol{r_{1}} &  \boldsymbol{r_{3}} & \ldots & 0 & 0\\
\vdots & \vdots & \vdots & \ddots & \vdots & \vdots \\
0 & 0 & 0 & \ldots & \boldsymbol{r_{3}}e^{ik\sqrt3 a} & \boldsymbol{r_{2}}\\
0 & 0 & 0 & \ldots & \boldsymbol{r_{1}} & \boldsymbol{r_{3}}
\end{pmatrix},
\end{align}
with $\mathbf{r_1}=a/\sqrt{3}(-\sqrt{3}/2,1/2)$; $\mathbf{r_2}=a/\sqrt{3}(\sqrt{3}/2,1/2)$; $\mathbf{r_3}=a/\sqrt{3}(0,-1)$.

\end{document}